\documentclass[jmp,preprint]{revtex4-1}
\usepackage{amsfonts}
\usepackage{mathrsfs}
\usepackage{graphicx}
\usepackage{amsmath}
\usepackage{amssymb}
\usepackage{amsthm}
\usepackage{hyperref}

\renewcommand\i{\mathrm{i}}
\renewcommand\Re{\hspace{0.2mm}\mathrm{Re}\hspace{0.2mm}}
\renewcommand\Im{\hspace{0.2mm}\mathrm{Im}\hspace{0.2mm}}
\begin{document}

\title[The Hahn Quantum System]
{The Hahn Quantum System}

\author{A. D. Alhaidari}
\email{haidari@sctp.org.sa}
\altaffiliation{Saudi Center for Theoretical Physics, P. O. Box 32741, Jeddah 21438, Saudi Arabia}

\author{Y.-T. Li}
\email{yutianlee@gmail.com}
\altaffiliation{
School of Science and Engineering, Chinese University of Hong Kong, Shenzhen, Guangdong, China}

\date{\today}

\begin{abstract}
Using a formulation of quantum mechanics based on the theory of orthogonal polynomials,
we introduce a four-parameter system associated with the Hahn and continuous Hahn polynomials.
The continuum energy scattering states are written in terms of the continuous Hahn polynomial
whose asymptotics give the scattering amplitude and phase shift.
On the other hand,
the finite number of discrete bound states are associated with the Hahn polynomial.
\end{abstract}

\pacs{03.65.Ca, 03.65.Nk, 03.65.Ge, 02.30.Gp}
\keywords{wavefunction, Hahn polynomials, asymptotics, phase shift, energy spectrum, resonances}

\maketitle

\section{Introduction}
\label{sec:1}

Using the well-established connection between scattering and the asymptotics of orthogonal polynomials
in the energy~\cite{case1974orthogonal,geronimo1980scattering,geronimo1980relation},
we introduced a formulation of quantum mechanics based on the theory of orthogonal polynomials
\cite{alhaidari2015formulation,alhaidari2015quantum,alhaidari2017wilson,alhaidari2017quantum}.
The traditional role of the potential function in describing the physical properties of the system
is taken up by a complete set of orthogonal polynomials.
All structural and dynamical features of the physical system are deduced from the properties of these polynomials.
For example, the bound state energies are obtained from the spectrum formula of the associated polynomial. Additionally, the energy density of states is obtained
from the distribution of the zeros of the polynomial for a large degree
(see, for example, the Appendix in Ref.~\cite{alhaidari2017density}).
In fact, these energy polynomials carry more information than the potential function.
For example, in three dimensional problems with spherical symmetry,
the polynomials already contain the angular momentum quantum number
whereas the potential function does not
(see, for example, the Coulomb problem and isotropic oscillator treated using this formulation in section 2 of Ref.~\cite{alhaidari2017quantum}).

The total wavefunction of a conservative quantum mechanical system at an energy $E$ is written as
$\Psi(t,x)=e^{-{\i}Et/\hbar}\psi(x,E)$ and the associated Hamiltonian acts on it as follows
$H\Psi=\i\hbar\frac{\partial}{\partial t}\Psi=E\Psi$.
In the proposed formulation,
the time-independent wavefunction of the system, $|\psi(x,E)\rangle$,
is treated as a local vector in an infinite dimensional space
and written in terms of its projections $\{f_n\}_{n=0}^\infty$  along ``local unit vectors'',
$\{|\phi_n(x)\rangle\}$.
That is, we write $|\psi(x,E)\rangle=\sum_{n}f_n|\phi_n(x)\rangle$.
In the language of calculus,
our terminology of ``local unit vectors'' means square integrable basis functions $\{\phi_n(x)\}$
and it is assumed that this sum is bounded.
Moreover, for a faithful representation of the physical system,
the basis $\{\phi_n(x)\}$ must form a complete set.
If the quantum system is parameterized by a set of real numbers $\{\mu\}$,
then the wavefunction projections would be written as the parameterized energy functions $\{f_n^{\mu}(E)\}_{n=0}^\infty$ and the state of the system at the energy $E$ is written as
\begin{equation}\label{eq:1.1}
\psi^{\mu}(x,E)=\sum_{n}f_n^{\mu}(E)\phi_n(x).
\end{equation}
If we write $f_n^{\mu}(E)=f_0^{\mu}(E)P_n^{\mu}(\varepsilon)$,
where $\varepsilon$ is some proper function of $E$ and $\{\mu\}$,
then $P_0^{\mu}(\varepsilon)=1$ and we have shown elsewhere~\cite{alhaidari2015quantum}
that completeness of the basis and normalization of the density of state make $\{P_n^{\mu}(\varepsilon)\}$ a complete set of orthogonal polynomials.
The corresponding weight function is $[f_0^{\mu}(E)]^2$
and the orthogonality relation reads as follows
\begin{equation}\label{eq:1.2}
\int\rho^\mu (\varepsilon)P_n^{\mu}(\varepsilon)P_m^{\mu}(\varepsilon)d\zeta(\varepsilon)=\delta_{n,m}\,,
\end{equation}
where $\rho^\mu (\varepsilon)=[f_0^{\mu}(E)]^2$
and $d\zeta(\varepsilon)$ is an appropriate energy integration measure.
Therefore, we can rewrite the wavefunction expansion (1) as follows
\begin{equation}\label{eq:1.3}
\psi^{\mu}(x,E)=\sqrt{\rho^{\mu}(\varepsilon)}\sum\nolimits_n P_n^{\mu}(\varepsilon)\phi_n(x).
\end{equation}
However,
physical requirements dictate that all physically relevant polynomials must have the following asymptotic ($n\to\infty$) behavior
\begin{equation}\label{eq:1.4}
P_n^{\mu}(\varepsilon)
\approx
n^{-\tau}A^{\mu}(\varepsilon)
\big\{\cos\big[n^\xi\theta(\varepsilon)+\varphi(\varepsilon)\log n+\delta^{\mu}(\varepsilon)\big]
+O\left(n^{-1}\right)\big\},
\end{equation}
where $\tau$ and $\xi$ are real positive constants that depend on the particular energy polynomial.
The studies in
\cite{case1974orthogonal,geronimo1980scattering,geronimo1980relation,alhaidari2015formulation,
alhaidari2015quantum,alhaidari2017wilson,alhaidari2017quantum}
show that $A^{\mu}(\varepsilon)$ is the scattering amplitude
and $\delta^{\mu}(\varepsilon)$ is the phase shift.
Bound states, if they exist, occur at discrete energies $\{E(\varepsilon_k)\}$
that make the scattering amplitude vanish, $A(\varepsilon_k)=0$.
The number of these bound states is either finite or infinite and we write the $k\textsuperscript{th}$ bound state as
\begin{equation}\label{eq:1.5}
\psi^{\mu}(x,E_k)=\sqrt{\omega^{\mu}(\varepsilon_k)}\sum_{n}Q_n^{\mu}(\varepsilon_k)\phi_n(x),
\end{equation}
where $\{Q_n^{\mu}(\varepsilon_k)\}$ are the discrete version of the polynomials $\{P_n^{\mu}(\varepsilon)\}$
and $\omega^{\mu}(\varepsilon_k)$ is the corresponding discrete weight function.
That is, $\sum_k \omega^{\mu}(\varepsilon_k)Q_n^{\mu}(\varepsilon_k)Q_m^{\mu}(\varepsilon_k)=\delta_{n,m}$.
If it happens that $A(\varepsilon_k)=0$ for complex $\{E(\varepsilon_k)\}$
then these correspond to resonances provided that the imaginary part of $E(\varepsilon_k)$ is negative
(clarification of this sign constraint is found after Eq.~(\ref{eq:2.2}) in the following section).

Using the polynomial formulation of quantum mechanics outlined above,
the authors in Ref.~\cite{alhaidari2015quantum} studied quantum systems corresponding to the two-parameter Meixner-Pollaczek polynomial and to the three-parameter continuous dual Hahn polynomial.
Special cases of these systems include, but not limited to, the Coulomb, oscillator and Morse problems.
Most notably though,
new systems that do not belong to the conventional class of exactly solvable problems were also found.
Their associated scattering phase shift and bound states energy spectra were obtained analytically.
In Ref.~\cite{alhaidari2017wilson} and using the same formulation,
we presented a four-parameter system associated with the Wilson polynomial
and its discrete version, the Racah polynomial.
Recently, we have shown for the first time that many of the well-known quantum mechanical systems are,
in fact, associated with the Wilson-Racah polynomial class~\cite{alhaidari2017quantum}.
These include, but not limited to, the P\"{o}schl-Teller, Scarf, Eckart,
and Rosen-Morse potentials (the trigonometric as well as the hyperbolic versions).

In the following section,
we introduce the four-parameter quantum systems associated with the continuous Hahn polynomial
and its discrete version, the Hahn polynomial.
We obtain the phase shift for the continuum scattering states and the bound/resonance energy spectrum for the discrete states.

\section{The Hahn quantum system}
\label{sec:2}

\begin{figure}[!htbp]
\includegraphics[width=0.5\textwidth]{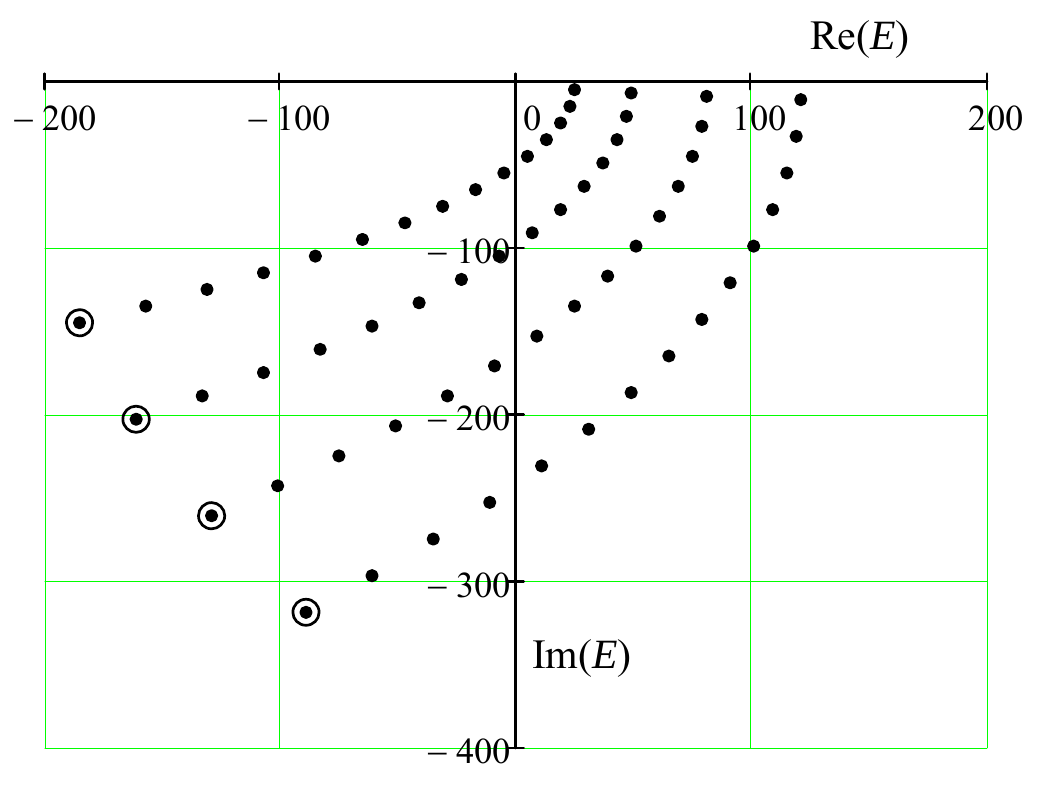}
\caption{Resonance energies (in units of $\lambda^2/2$)
associated with the spectrum formula of Eq.~(\ref{eq:2.2})
in the complex energy plane for $\mu=-14.5$ and several values of $a=-\{5,7,9,11\}$.
The circled resonances correspond to $k=0$ where $k=0,1,\ldots,14$.
Resonances with $k<a-\mu$ are ``\emph{bound states embedded resonances}''.}
\label{fig:1}
\end{figure}

For this system, the expansion coefficients of the continuous energy wavefunction in Eq.~(\ref{eq:1.3})
are the four-parameter continuous Hahn polynomials.
The normalized version of this polynomial is given in Appendix~\ref{sec:A} by Eq.~(\ref{eq:A.1})
and the corresponding normalized weight function is given by Eq.~(\ref{eq:A.2}).
Moreover, the asymptotic formula for this polynomial is derived in Appendix~\ref{sec:B} and given by (\ref{eq:B.6}).
As a physical example, we choose $\nu=\mu$, $b=-a$ and $z=\kappa/\lambda$,
where $\kappa$ is the wavenumber where $E=\frac12\kappa^2$,
and $\lambda^{-1}$ is a length scale parameter in the atomic units $\hbar=m=1$.
The scattering phase shift given by Eq.~(\ref{eq:A.5}) becomes
\begin{equation}\label{eq:2.1}
\delta(E)=-2\arg \Gamma\big[\mu+{\i}(a+\kappa/\lambda)\big],
\end{equation}
where the parameter $\mu$ is positive.
For $\mu<0$, the spectrum formula (\ref{eq:A.6}) gives
\begin{equation}\label{eq:2.2}
E_k=-\frac{\lambda^2}{2}(k+\mu+{\i}a)^2,
\end{equation}
where $k=0,1,\ldots,N$ and $N$ is the largest integer less than or equal to $-\mu$.
These values are real only if $a=0$ otherwise they are complex.
Thus, we conclude that bound states exist only if $a=0$
where the finite energy spectrum is $E_k=-\frac12\lambda^2(k+\mu)^2$
and the corresponding bound-sate wavefunctions are written as Eq. (\ref{eq:1.5})
with the discrete Hahn polynomials as expansion coefficients.
Now, the time phase factor $e^{-{\i}Et}$ in the total wavefunction $\Psi(t,x)$
implies that if $E$ is complex then its imaginary part must be negative
so that the corresponding state decays in time indicating resonance,
otherwise the state will grow unphysically with time.
Therefore, $a$ must be negative and Fig.~\ref{fig:1} shows the location of these resonance energies
in the lower half of the complex energy plane for a given $\mu$ and several values of $a$.
Resonances with $k<a-\mu$ are located in the third quadrant of the complex energy plane
(i.e., with negative real energy part).
Sometimes, these are referred to as ``\emph{bound states embedded resonances}''~\cite{alhaidari2005unified}.

\begin{figure}[!htbp]
\includegraphics[width=0.5\textwidth]{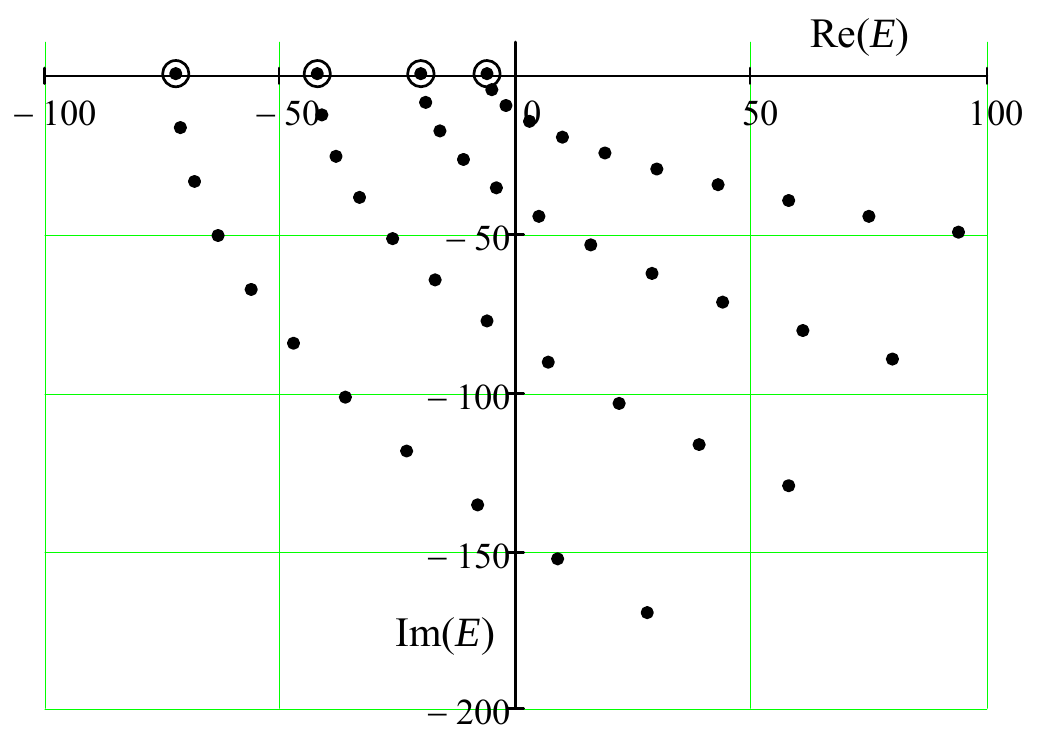}
\caption{Resonance energies (in units of $\lambda^2/2$)
associated with the spectrum formula of Eq.~(\ref{eq:2.4})
in the complex energy plane for $V=7.5$ (in units of $\lambda^2/2$)
and several values of $a=-\{10,12,14,16\}$.
The circled resonances correspond to $k=0$
and we have shown only the lower part of the resonance spectrum for $k=0,1,\ldots,10$.
Resonances with $k<-(a+2V/\lambda^2)$ are ``\emph{bound states embedded resonances}''.}
\label{fig:2}
\end{figure}

We construct the second example by making a different selection of polynomial parameters.
Let us choose $\mu=\nu=\kappa/\lambda$ and $z=2V/\lambda^2$
with $V$ being a real parameter of inverse squared length dimension.
The scattering phase shift is parameterized by $V$, $a$ and $b$ and reads as follows
\begin{equation}\label{eq:2.3}
\delta(E)=-\arg\Gamma\left[\frac{2}{\lambda^2}(E+{\i}V)+{\i}a\right]
-\arg\Gamma\left[\frac{2}{\lambda^2}(E+{\i}V)-{\i}b\right].
\end{equation}
The spectrum formula (\ref{eq:A.6}) gives the following discrete energies
\begin{equation}\label{eq:2.4}
E_k=\frac{\lambda^2}{2}\left[k+{\i}\left(a+\frac{2V}{\lambda^2}\right)\right]^2.
\end{equation}
Therefore, bound states exist only if $a=-2V/\lambda^2$
where the energy spectrum becomes simply $E_k=\frac{1}{2}\lambda^2k^2$ and it is of an infinite size
(i.e., $N\to\infty$).
On the other hand, the energies in (\ref{eq:2.4}) with $a<-2V/\lambda^2$ correspond to resonances
for all $k>-(a+2V/\lambda^2)$
and they are located in the fourth quadrant in the complex energy plane.
Figure~\ref{fig:2} shows the location of these resonance energies in the lower half of the complex energy plane
for a given $V$ and several values of $a$.

\begin{figure}[!htbp]
\includegraphics[width=0.5\textwidth]{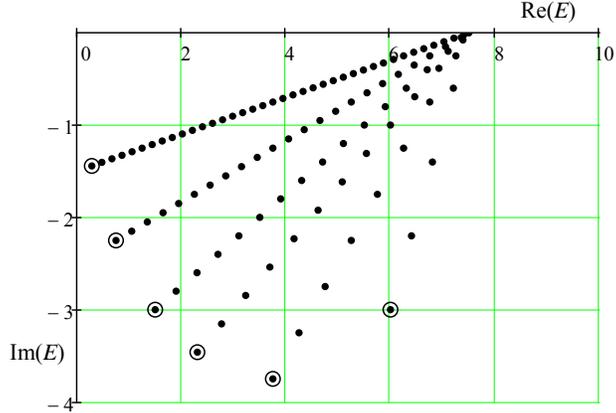}
\caption{Resonance energies (in units of $\lambda^2$)
associated with the spectrum formula of Eq.~(\ref{eq:2.6})
in the complex energy plane for $a=-7.5$
and several values of $\gamma=-\{\frac12,1,\frac32,2,3,5\}$.
The circled resonances correspond to $k=0$ where $k=0,1,\ldots,a\gamma$.
The linear resonance chains intersect the real energy line at $E=-\lambda^2a$.}
\label{fig:3}
\end{figure}

The third and final example is associated with the following parametrization:
$z=E/\lambda^2$, $\mu=\gamma E/\lambda^2$ where $\gamma$ is a dimensionless real parameter.
The scattering phase shift for $b=-a$ becomes
\begin{equation}\label{eq:2.5}
\delta(E)=-\arg\Gamma\left[\frac{E}{\lambda^2}(\gamma+{\i})+{\i}a\right]
-\arg\Gamma\left[\nu+\frac{\i}{\lambda^2}(E+a\lambda^2)\right].
\end{equation}
The spectrum formula (\ref{eq:A.6}) results only in resonance energies as follows
\begin{equation}\label{eq:2.6}
\frac{E_k}{\lambda^2}=\frac{-(a+k\gamma)+{\i}(k-a\gamma)}{1+\gamma^2},
\end{equation}
where $k=0,1\ldots,N$ and $N$ is the largest integer less than or equal to $\gamma a$ where $\gamma a>0$.
These are displayed in Fig.~\ref{fig:3} for a given $a$ and several values of $\gamma$.

It should be noted that we have obtained all properties of the physical system,
including the scattering phase shift, bound states and resonance energies,
without specifying any basis set $\{\phi_n(x)\}$.
Choosing a basis will fix the physical configuration of the problem that corresponds to these physical properties.
In other words, there are possibly many physical configurations with the same physical features.
To understand this, we make a very simple analogy to standard vector quantities in physics
such as the force and electric field, etc.
We can write these vector quantities in, say, the Cartesian coordinates
with basis unit vectors $\{\hat{x},\hat{y},\hat{z}\}$ as $\vec{F}=f_x\hat{x}+f_y\hat{y}+f_z\hat{z}$.
All physical information about the quantity $\vec{F}$ are contained in its components $\{f_x,f_y,f_z\}$
whereas the basis unit vectors $\{\hat{x},\hat{y},\hat{z}\}$ are essentially dummy;
they are only required to be complete so that a faithful representation of $\vec{F}$ is obtained.
In fact, we can still write the same physical quantity in another coordinate system,
such as spherical coordinates with basis unit vectors $\{\hat{r},\hat{\theta},\hat{\varphi}\}$,
as $\vec{F}=f_r\hat{r}+f_\theta\hat{\theta}+f_\varphi\hat{\varphi}$.
However, the new components $\{f_r,f_\theta,f_\varphi\}$ are different from the Cartesian components
$\{f_x,f_y,f_z\}$, but they contain the same physical information.
On the other hand, if we keep the same components in the new basis by writing $\vec{F}=f_x\hat{r}+f_y\hat{\theta}+f_z\hat{\varphi}$
then the result will be a physically different $\vec{F}$.
In this simple analogue and by comparison to Eq. (\ref{eq:1.1}),
the objects $\vec{F}$, $\{f_x,f_y,f_z\}$, and $\{\hat{x},\hat{y},\hat{z}\}$
play the same role as $\psi^{\mu}(x,E)$, $\{P_n^{\mu}(\varepsilon)\}$ and $\{\phi_n(x)\}$, respectively.
Thus, keeping the same polynomials while changing the basis results in a different physical setting.
However, unlike the basis unit vectors $\{\hat{x},\hat{y},\hat{z}\}$
there is an additional restriction on the choice of basis $\{\phi_n(x)\}$.
It goes as follows:
Since the orthogonal polynomials satisfy a three-term recursion relation
(e.g., Eq. (\ref{eq:A.3}) in Appendix~\ref{sec:A}),
then the basis must produce a tridiagonal matrix representation for the wave operator.
Because only then will the matrix wave equation become equivalent to the three-term recursion relation.
Technical details where this matter is analyzed is given in Ref.~\cite{alhaidari2017establishing}.
However, for ease of reference, we give a general outline here and as follows.
The wave equation in configuration space is $[H(x)-E]|\psi(x)\rangle=0$,
where $H(x)$ is the Hamiltonian operator. Using the expansion of the wave function in Eq. (\ref{eq:1.3}),
we can rewrite this as $\sqrt{\rho^\mu}\sum_{n}P_n^{\mu}J(x)|\phi_n(x)\rangle=0$,
where $J(x)$ is the wave operator $H(x)-E$.
Projecting from left on this equation by $\langle \phi_m(x)|$
and writing the matrix elements of the wave operator as $J_{mn}=\langle \phi_m(x)|J(x)|\phi_n(x)\rangle$,
we obtain the matrix wave equation $\sum_n J_{mn}P_n^{\mu}=0$.
If this is to be equivalent to the recursion relation (\ref{eq:A.3}),
then the matrix representation of the wave operator $J$ in the basis $\{\phi_n\}$
must be tridiagonal and symmetric giving
\begin{equation}\label{eq:2.7}
J_{n,n}P_{n}^{\mu}(\varepsilon)
+J_{n,n-1}P_{n-1}^{\mu}(\varepsilon)
+J_{n,n+1}P_{n+1}^{\mu}(\varepsilon)=0.
\end{equation}
Note that the symmetry requirement is guaranteed by the Hermiticity of the Hamiltonian.
In the following section,
we consider specific physical configurations associated with the continuous Hahn polynomial
and reconstruct the corresponding potential function.

\section{Potential functions associated with the Hahn system}
\label{sec:3}

\begin{figure}[!htbp]
\includegraphics[width=0.6\textwidth]{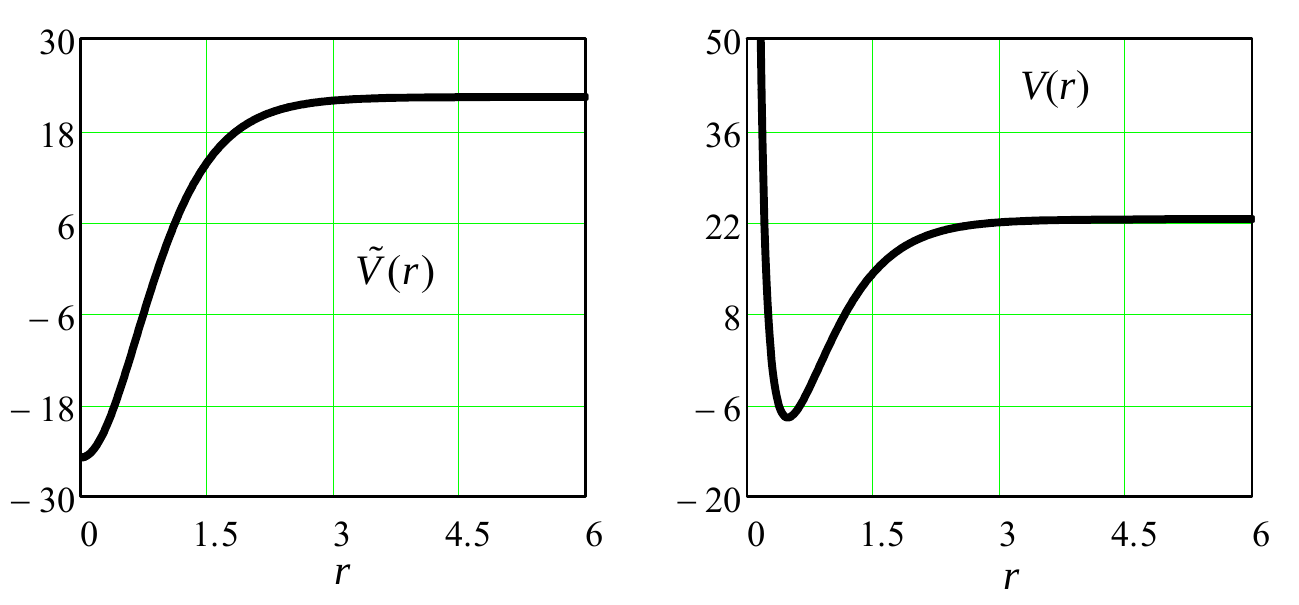}
\caption{The potential functions $\widetilde{V}(r)$ and $V(r)$
for the first example in Sec.~\ref{sec:3},
which is associated with the Jacobi basis (\ref{eq:3.1}).
We took the physical parameters $\{\lambda,\alpha,\beta,\mu,\nu,a\}=\{1,5,2,3,4,2\}$.}
\label{fig:4}
\end{figure}

In this section, we construct the potential function and wavefunction
associated with four examples of Hahn quantum mechanical systems.
We start by choosing a complete set of square integrable basis functions.
For the first two examples,
we consider the following elements of the Jacobi basis
\begin{equation}\label{eq:3.1}
\phi_n(x)
=A_n(1-y)^{\sigma}(1+y)^{\tau} P_{n}^{(\alpha,\beta)}(y),
\end{equation}
where $P_{n}^{(\alpha,\beta)}(y)$ is the Jacobi polynomial of degree $n$ in $y$ and $-1\leq y(x)\leq +1$.
$A_n$ is the normalization constant whereas the parameters $\alpha$ and $\beta$ are greater than $-1$.
First, we consider the problem in three dimensions with spherical symmetry and zero angular momentum
and take $y(r)=2\tanh^2(\lambda r)-1$, where $r$ is the radial coordinate
and $\lambda^{-1}$ is a length scale parameter.
If we also choose the basis (\ref{eq:3.1}) to be orthonormal
(i.e., $\langle \phi_n|\phi_m\rangle=\delta_{nm}$) with respect to the integration measure $\lambda dr$
then we must take $2\sigma=\alpha+1$, $2\tau=\beta+\frac12$
and choose $A_n=\sqrt{\frac{2n+\alpha+\beta+1}{2^{\alpha+\beta+\frac12}}
\frac{\Gamma(n+1)\Gamma(n+\alpha+\beta+1)}{\Gamma(n+\beta+1)\Gamma(n+\alpha+1)}}$.
We write the Hamiltonian operator as $H=T+V(r)$,
where $T$ is the kinetic energy operator $-\frac12\frac{d^2}{r^2}$
and $V(r)$ is the potential function.
Using the differential equation, recursion relation, and orthogonality of the Jacobi polynomials,
the work in section III.B.1 of Ref.~\cite{alhaidari2017solution} shows that
if we write $V(r)=\frac{V_0}{\sinh^2(\lambda r)}+\widetilde{V}(r)$,
where $V_0=(\beta^2-\frac14)\frac{\lambda^2}{2}$,
then the matrix representation of $T+\frac{V_0}{\sinh^2(\lambda r)}$,
which we refer to as the reference Hamiltonian $H_0$,
in the basis (\ref{eq:3.1}) is tridiagonal and symmetric as follows
\begin{align}
\frac{1}{\lambda^2}\left(H_0\right)_{n,m}
=&\left\{-\frac{2n(n+\beta)}{2n+\alpha+\beta}-\frac{(\alpha+1)^2}{2}
+\left[\left(n+\frac{\alpha+\beta}{2}+1\right)^2-\frac{1}{16}\right](1-C_n)
\right\}\delta_{n,m}
\nonumber
\\
&-\left[\left(n+\frac{\alpha+\beta}{2}\right)^2-\frac{1}{16}\right]\
D_{n-1}\delta_{n,m+1}
-\left[\left(n+\frac{\alpha+\beta}{2}+1\right)^2-\frac{1}{16}\right]\
D_{n}\delta_{n,m-1}\, ,
\label{eq:3.2}
\end{align}
where $C_n=\frac{\beta^2-\alpha^2}{(2n+\alpha+\beta)(2n+\alpha+\beta+2)}$,
$D_n=\frac{2}{2n+\alpha+\beta+2}
\sqrt{\frac{(n+1)(n+\alpha+1)(n+\beta+1)(n+\alpha+\beta+1)}{(2n+\alpha+\beta+1)(2n+\alpha+\beta+3)}}$.
Now, the matrix wave equation in the orthonormal basis (\ref{eq:3.1})
is $J_{nm}=H_{nm}-E\delta_{nm}$ making Eq.~(\ref{eq:2.7}) reads as follows
\begin{equation}\label{eq:3.3}
E P_{n}^{\mu}(\varepsilon)
=H_{n,n}P_{n}^{\mu}(\varepsilon)
+H_{n,n-1}P_{n-1}^{\mu}(\varepsilon)
+H_{n,n+1}P_{n+1}^{\mu}(\varepsilon).
\end{equation}
Comparing this equation with the three-term recursion relation of the continuous Hahn polynomial (\ref{eq:A.3})  and taking $E=\lambda^2z$,
we obtain the following elements of the tridiagonal Hamiltonian matrix
\begin{subequations}\label{eq:3.4}
\begin{align}
&\frac{1}{\lambda^2}H_{n,n}
=-a+\frac{(a+b)/2}{2n+2\mu+2\nu-1}
\left[
\frac{(n+2\mu)(n+2\mu+2\nu-1)}{n+\mu+\nu}
+\frac{n(n+2\nu-1)}{n+\mu+\nu-1}
\right],
\\
&\frac{1}{\lambda^2}H_{n,n+1}
=\frac{1}{\lambda^2}H_{n+1,n}
\\
=&\frac{1/2}{n+\mu+\nu}
\sqrt{
\frac{(n+1)(n+2\mu)(n+2\nu)(n+2\mu+2\nu-1)[(n+\mu+\nu)^2+(a+b)^2]}
{(2n+2\mu+2\nu-1)(2n+2\mu+2\nu+1)}
}.
\nonumber
\end{align}
\end{subequations}
Thus, the matrix elements of $H_0$ and $H$ are now given in terms of the parameter set
$\{\alpha,\beta,\mu,\nu,a,b,\lambda\}$.
Consequently, the matrix elements of the potential $\widetilde{V}(r)$
in the basis (\ref{eq:3.1}) are obatined as
$\widetilde{V}_{nm}=H_{nm}-(H_0)_{nm}$.
Using one of four methods developed in Ref.~\cite{alhaidari2017establishing},
we can obtain a very good approximation of the potential function $\widetilde{V}(r)$
using only its matrix elements and the basis (\ref{eq:3.1}) in which they are represented.
For example, the second method established in Section 3.2 of~\cite{alhaidari2017establishing} gives
\begin{equation}\label{eq:3.5}
\widetilde{V}(x)\cong\sum_{m=0}^{M-1} \frac{\phi_m(x)}{\phi_0(x)}\widetilde{V}_{m,0}\,,
\end{equation}
where $M$ is some large enough integer.
Rescaling the energy and potential parameters by $\lambda^2$ and taking $b=a$,
Eq.~(\ref{eq:3.5}) gives the potential functions $\widetilde{V}(r)$ and $V(r)$
show in Fig.~\ref{fig:4} for a given set of parameters $\{\alpha,\beta,\mu,\nu,a\}$.

\begin{figure}[!htbp]
\includegraphics[width=0.6\textwidth]{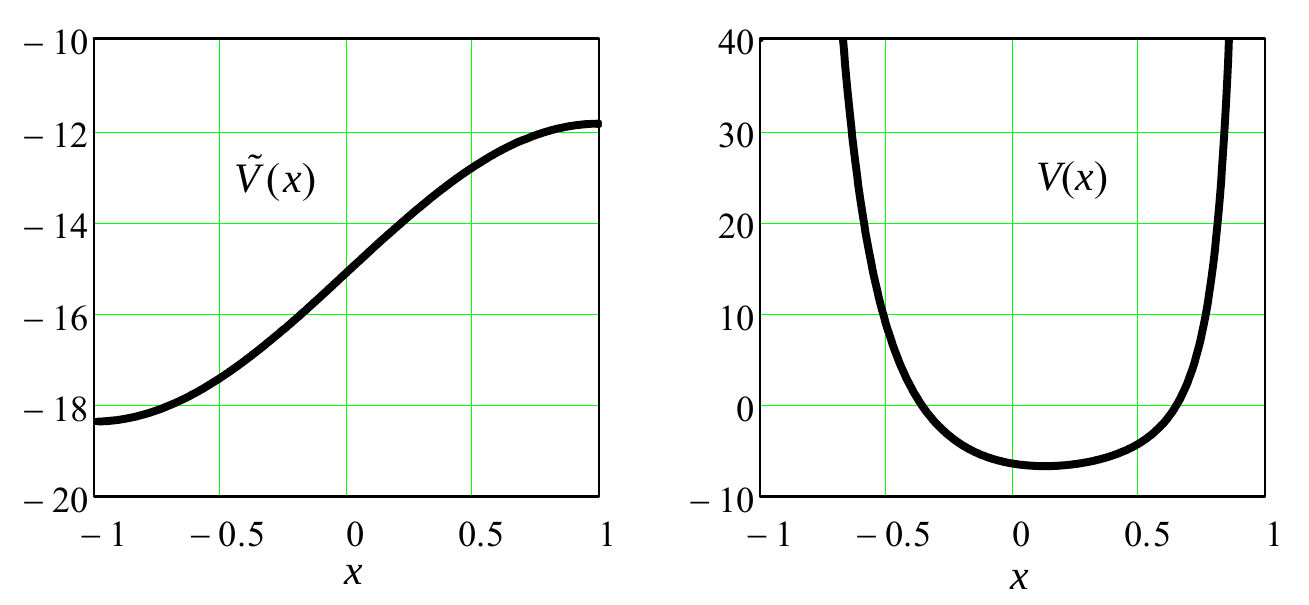}
\caption{The potential functions $\widetilde{V}(x)$ and $V(x)$
for the second example in Sec.~\ref{sec:3},
which is associated with the Jacobi basis (\ref{eq:3.1}).
We took the physical parameters $\{L,\alpha,\beta,\mu,\nu,a\}=\{2,1.5,3.5,3,4,2.5\}$.}
\label{fig:5}
\end{figure}

For the second example,
we repeat the same analysis in the basis (\ref{eq:3.1})
but in one dimension with $y(x)=\sin(\pi x/L)$ where $-L/2\leq x\leq +L/2$.
Choosing $2\sigma=\alpha+\frac12$ and $2\tau=\beta+\frac12$ will make the basis orthonormal
if we also take the normalization constant as
$A_n=\sqrt{\frac{2n+\alpha+\beta+1}{2^{\alpha+\beta+1}}
\frac{\Gamma(n+1)\Gamma(n+\alpha+\beta+1)}{\Gamma(n+\beta+1)\Gamma(n+\alpha+1)}}$
with $\lambda=\pi/L$.
The work in section III.A.1 of Ref.~\cite{alhaidari2017solution} shows that
if we write
$V(x)=\frac{V_{+}-V_{-}\sin(\pi x/L)}{\cos^2(\pi x/L)}+\widetilde{V}(x)$,
where $V_{+}=(\alpha^2+\beta^2-\frac12)\frac{\lambda^2}{4}$
and $V_{-}=(\beta^2-\alpha^2)\frac{\lambda^2}{4}$,
then the matrix representation of the Hamiltonian
$H_0=-\frac12\frac{d^2}{dx^2}+\frac{V_{+}-V_{-}\sin(\pi x/L)}{\cos^2(\pi x/L)}$
in the basis (\ref{eq:3.1}) becomes
\begin{equation}\label{eq:3.6}
(H_0)_{nm}=\frac{\lambda^2}{2}\left(n+\frac{\alpha+\beta+1}{2}\right)^2\delta_{nm}\, .
\end{equation}
The total Hamiltonian matrix is still given by Eq. (\ref{eq:3.4}) above.
Hence, the matrix elements of the potential function $\widetilde{V}(x)$ are now easily obtained
as $\widetilde{V}_{nm}=H_{nm}-(H_0)_{nm}$.
Using these $\{\widetilde{V}_{nm}\}$ and the basis elements (\ref{eq:3.1})
together with the choice $b=-a$,
the second method of \cite{alhaidari2017establishing}
as depicted by (\ref{eq:3.5})
gives the potential functions $\widetilde{V}(x)$ and $V(x)$ shown in Fig.~\ref{fig:5}
for a given set of polynomial parameters $\{\alpha,\beta,\mu,\nu,a\}$.

\begin{figure}[!htbp]
\includegraphics[width=0.6\textwidth]{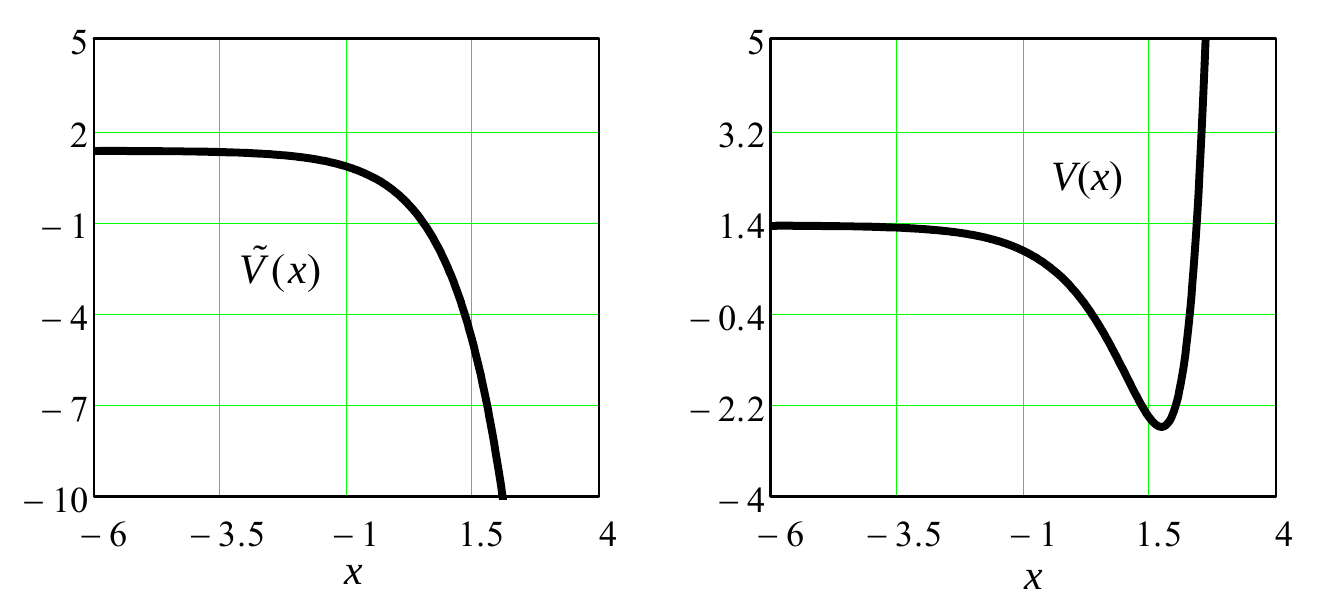}
\caption{The potential functions $\widetilde{V}(x)$ and $V(x)$
for the third example in Sec.~\ref{sec:3},
which is associated with the Jacobi basis (\ref{eq:3.7}).
We took the physical parameters $\{\lambda,\beta,\mu,\nu,a\}=\{1,2.7,3,4,2.5\}$.}
\label{fig:6}
\end{figure}

For the next two examples, we consider the following Laguerre basis
\begin{equation}\label{eq:3.7}
\phi_n(x)=A_n y^{\alpha}e^{-y/2}L_n^{\beta}(y),
\end{equation}
where $L_n^{\beta}(y)$ is the Laguerre polynomial of degree $n$ in $y$ and $y(x)\geq 0$.
$A_n$ is the normalization constant whereas the dimensionless real parameter $\beta$ is greater than $-1$.
We start by considering the one dimensional problem with $y(x)=e^{\lambda x}$ and $-\infty<x<+\infty$.
If we also like to work in an orthonormal basis set then we must choose $2\alpha=\beta+1$ and take
$A_n=\sqrt{\frac{\Gamma(n+1)}{\Gamma(n+\beta+1)}}$.
In section II.B.1 of Ref.~\cite{alhaidari2017solution},
we show that if we write
$V(x)=\frac{\lambda^2}{8}e^{2\lambda x}+\widetilde{V}(x)$,
then the matrix representation of the reference Hamiltonian
$H_0=-\frac12\frac{d^2}{dx^2}+\frac{\lambda^2}{8}e^{2\lambda x}$
in the basis (\ref{eq:3.1}) is tridiagonal and symmetric as follows
\begin{align}
\label{eq:3.8}
&\frac{2}{\lambda^2}(H_0)_{nm}
=\left[(2n+\beta+1)\left(n+\frac{\beta}{2}+1\right)-n-\frac{(\beta+1)^2}{4}\right]
\delta_{n,m}
\\
-&\left(n+\frac{\beta}{2}\right)\sqrt{n(n+\beta)}\delta_{n,m+1}
-\left(n+\frac{\beta}{2}+1\right)\sqrt{(n+1)(n+\beta+1)}\delta_{n,m-1}\,.
\nonumber
\end{align}
The total Hamiltonian matrix is still given by Eq.~(\ref{eq:3.4}).
Using that and (\ref{eq:3.8}),
we obtain the matrix elements of the potential $\widetilde{V}(x)$
as $\widetilde{V}_{nm}=H_{nm}-(H_0)_{nm}$.
With $E=\lambda^2z$ and $b=-a$,
we obtain the potential functions $\widetilde{V}(x)$ and $V(x)$
shown in Fig.~\ref{fig:6} using the second method in Section 3.2 of Ref.~\cite{alhaidari2017establishing}
as portrayed by Eq.~(\ref{eq:3.5}) and for a given set of values of the parameters
$\{\lambda,\beta,\mu,\nu,a\}$.

\begin{figure}[!htbp]
\includegraphics[width=0.6\textwidth]{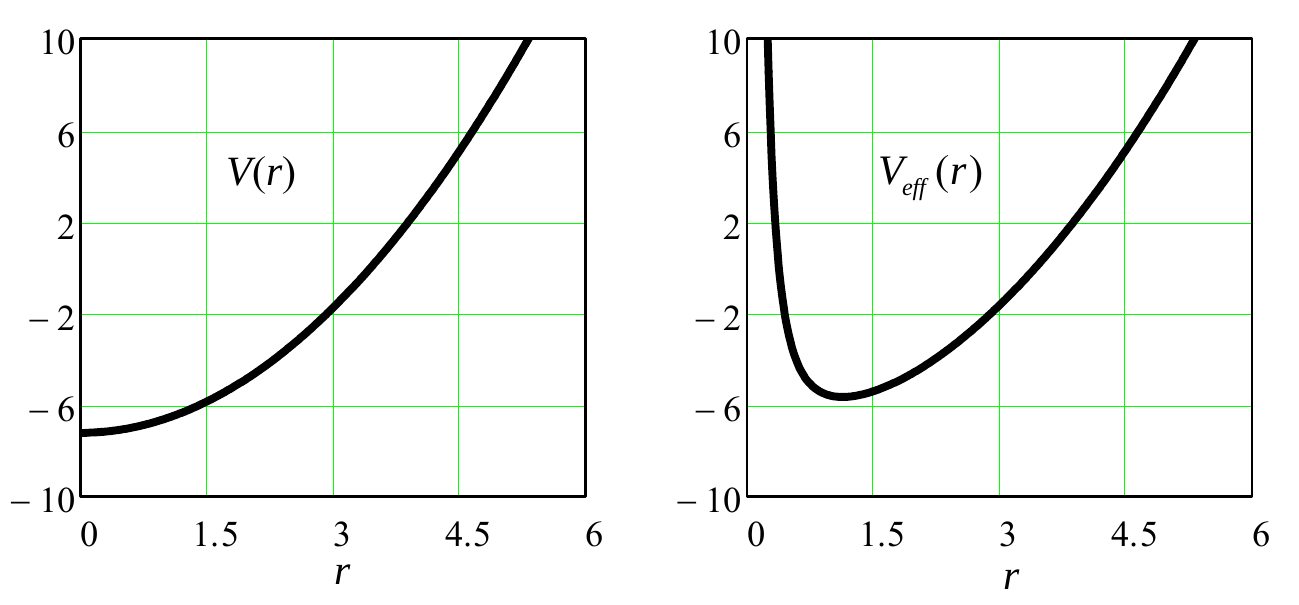}
\caption{The potential functions $V(r)$ and $V_{\textit{eff}}(r)=\frac{\ell(\ell+1)}{2r^2}+V(r)$
for the fourth example in Sec.~\ref{sec:3},
which is associated with the Jacobi basis (\ref{eq:3.7}).
We took the physical parameters $\{\lambda,\ell,\mu,\nu,a\}=\{2,1,3,4,2.5\}$.}
\label{fig:7}
\end{figure}

The fourth and final example is in the Laguerre basis (\ref{eq:3.7})
and in three dimensions with spherical symmetry and angular momentum quantum number $\ell=0,1,\ldots$.
We take $y(r)=(\frac12\lambda r)^2$, $2\alpha=\beta+\frac12$ and $A_n=\sqrt{\frac{\Gamma(n+1)}{\Gamma(n+\beta+1)}}$ resulting in an orthonormal basis.
In section II.A.1 of Ref.~\cite{alhaidari2017solution},
we show that if we choose $\beta=\ell+\frac12$
then the matrix representation of the kinetic energy operator
$-\frac12\frac{d^2}{dr^2}+\frac{\ell(\ell+1)}{2r^2}$
becomes tridiagonal and symmetric as follows
\begin{equation}\label{eq:3.9}
\frac{2}{\lambda^2}T_{n,m}
=(2n+\beta+1)\delta_{n,m}
+\left[
\sqrt{n(n+\beta)}\delta_{n,m+1}
+\sqrt{(n+1)(n+\beta+1)}\delta_{n,m-1}
\right].
\end{equation}
Using this and the total Hamiltonian matrix (\ref{eq:3.4}),
we obtain the matrix elements of the potential $V(r)$ as $V_{nm}=H_{nm}-T_{nm}$.
With $E=\lambda^2z$ and $b=a$,
we obtain the potential function $V(r)$ shown in Fig.~\ref{fig:7}
using the second method in Ref.~\cite{alhaidari2017establishing}
as represented by Eq.~(\ref{eq:3.5})
and for a given set of parameters $\{\ell,\mu,\nu,a\}$.

It is our observation that in all four examples above the sought after potential function
$\widetilde{V}$ ($V$ in the last one) is found to be a linear function of $y$.
That is, $\widetilde{V}(x)=\widetilde{V_0}+\widetilde{V}_1y(x)$,
where $\widetilde{V_0}$ and $\widetilde{V_1}$ are real constants that depend on the parameters
$\{\alpha,\beta,\mu,\nu,a,b,\lambda,\ell\}$.
This linear behavior could be explained as follows.
The matrix representation of the potential function $\widetilde{V}(x)$ is
\begin{equation}\label{eq:3.10}
\widetilde{V}_{nm}
=\langle\phi_n|\widetilde{V}|\phi_m\rangle
=\lambda\int\phi_n(x)\widetilde{V}(x)\phi_m(x)\frac{dy}{y'}
=A_nA_m\int\rho(y)p_n(y)\widetilde{V}(x)p_m(y)dy,
\end{equation}
where $y'=\frac{dy}{dx}$, $p_n(y)$ is the Jacobi or Laguerre polynomial
and $\rho(y)$ is the associated weight function.
Now, $H$ and $H_0$ are tridiagonal matrices; thus, so is $\{\widetilde{V}_{nm}\}$.
The three-term recursion relation and orthogonality of the Jacobi and Laguerre polynomials dictate that
for this to happen $\widetilde{V}(x)$ must be a linear function of $y$.
Therefore, we end up with the following:

\begin{enumerate}
\item In the first example,
$\widetilde{V}(r)=\widetilde{V}_0+\widetilde{V}_1[2\tanh^2(\lambda r)-1]$
making
$V(r)=\frac{V_0}{\sinh^2(\lambda r)}+\widetilde{V}_1[2\tanh^2(\lambda r)-1]+\widetilde{V}_0$
and to force the potential to vanish at infinity we must choose
$\widetilde{V}_0=-\widetilde{V}_1$ giving finally
$V(r)=\frac{V_0}{\sinh^2(\lambda r)}-\frac{2\widetilde{V}_1}{\cosh^2(\lambda r)}$,
which is the hyperbolic P\"{o}schl-Teller potential.

\item In the second example,
$\widetilde{V}(x)=\widetilde{V}_0+\widetilde{V}_1\sin(\pi x/L)$
making
$V(x)=\frac{V_{+}-V_{-}\sin(\pi x/L)}{\cos^2(\pi x/L)}+\widetilde{V}_1\sin(\pi x/L)+\widetilde{V}_0$,
which is a generalization of the trigonometric Scarf potential~\cite{al2017energy}.

\item In the third example,
$\widetilde{V}(x)=\widetilde{V}_0+\widetilde{V}_1e^{\lambda x}$
making
$V(x)=\frac{\lambda^2}{8}e^{2\lambda x}+\widetilde{V}_1e^{\lambda x}+\widetilde{V}_0$,
and to force the potential to vanish at $x=-\infty$
we must chose $\widetilde{V}_0=0$ resulting in
the one-dimensional Morse potential.

\item In the forth example,
$V(r)=\widetilde{V}(r)=\widetilde{V}_0+\widetilde{V}_1r^2$,
which is the three-dimensional isotropic oscillator.
\end{enumerate}

\section{Conclusion}
\label{sec:4}

Using a formulation of quantum mechanics based on orthogonal polynomials,
we introduced a new four-parameter quantum system whose scattering states are associated with
the continuous Hahn polynomial and bound states are associated with its discrete version, the Hahn polynomial. These polynomials constitute the expansion coefficients of the wavefunction in a complete set of
square integrable basis elements that produce a tridiagonal matrix representation for the wave operator. Depending on the values of the physical parameters,
the system consists of either continuous energy scattering states
or a finite number of discrete energy bound/resonance states.
The scattering phase shift and energy spectrum were obtained analytically.
To establish correspondence with the standard formulation of quantum mechanics,
we also obtained the potential functions associated with several physical configurations.

\section*{Acknowledgements}
ADH appreciates the support by the Saudi Center for Theoretical Physics (SCTP) during the progress of this work. YTL would like to thank Liu Bie Ju Centre for Mathematical Sciences at City University of Hong
Kong for its hospitality.

\appendix

\section{The Hahn and continuous Hahn polynomials}
\label{sec:A}

In theoretical physics,
we usually associate the coefficients in the recursion relation of relevant orthogonal polynomials
with the real elements of tridiagonal Hamiltonian matrices.
Due to the Hermiticity of these matrices, they are symmetric.
Therefore, we choose to work with the normalized version of the polynomials
where the corresponding three-term recursion relation becomes symmetric.
Now, the normalized version of the continuous Hahn polynomial reads as follows
(see pages 200-204 in Ref.~\cite{koekoek2010hypergeometric})
\begin{equation}\label{eq:A.1}
\begin{aligned}
P_n^{\mu}(z;\nu;a,b)
=&{\i}^n\sqrt{\left(\frac{2n+2\mu+2\nu-1}{2\mu+2\nu-1}\right)
\frac{(2\mu)_n(\mu+\nu+{\i}a+{\i}b)_n}{(2\nu)_n(\mu+\nu-{\i}a-{\i}b)_n}\frac{(2\mu+2\nu-1)_n}{n!}}
\\
&\times{}_{3}F_{2}
\left(\left.
{-n,n+2\mu+2\nu-1,\mu+{\i}(z+a)}\atop
{2\mu, \mu+\nu+{\i}(a+b)}
\right|1\right),
\end{aligned}
\end{equation}
where $\mu$ and $\nu$ are positive and $z\in\mathbb{R}$.
The normalized weight function is
\begin{equation}\label{eq:A.2}
\rho^\mu(z;\nu;a,b)=\frac{1}{2\pi}
\frac{\Gamma(2\mu+2\nu)\big|\Gamma[\mu+{\i}(z+a)]\Gamma[\nu+{\i}(z-b)]\big|^2}
{\Gamma(2\mu)\Gamma(2\nu)\big|\Gamma[\mu+\nu+{\i}(a+b)]\big|^2}.
\end{equation}
Thus
$\int_{-\infty}^{+\infty}\rho^\mu(z;\nu;a,b)P_n^{\mu}(z;\nu;a,b)P_m^{\mu}(z;\nu;a,b)dz=\delta_{nm}$.
It also satisfies the following symmetric three-term recursion relation
\begin{equation}\label{eq:A.3}
\footnotesize{
\begin{aligned}
&2(z+a)P_n^{\mu}=
\frac{a+b}{2n+2\mu+2\nu-1}
\left[\frac{(n+2\mu)(n+2\mu+2\nu-1)}{n+\mu+\nu}+\frac{n(n+2\nu-1)}{n+\mu+\nu-1}\right]P_n^{\mu}
\\
&+\frac{1}{n+\mu+\nu}
\sqrt{\frac{(n+1)(n+2\mu)(n+2\nu)(n+2\mu+2\nu-1)[(n+\mu+\nu)^2+(a+b)^2]}
{(2n+2\mu+2\nu-1)(2n+2\mu+2\nu+1)}}P_{n+1}^{\mu}
\\
&+\frac{1}{n+\mu+\nu-1}
\sqrt{\frac{n(n+2\mu-1)(n+2\nu-1)(n+2\mu+2\nu-2)[(n+\mu+\nu-1)^2+(a+b)^2]}
{(2n+2\mu+2\nu-3)(2n+2\mu+2\nu-1)}}P_{n-1}^{\mu}.
\end{aligned}}
\end{equation}
The asymptotics ($n\to\infty$) of the continuous Hahn polynomial
is derived in Appendix~\ref{sec:B} and shown as formula (\ref{eq:B.6}).
Comparing that with the general formula of Eq.~(\ref{eq:1.4}), we obtain the scattering amplitude and phase shift as follows
\begin{equation}\label{eq:A.4}
A^{\mu}(\varepsilon)
=2\sqrt{\frac{2\Gamma(2\mu)\Gamma(2\nu)}{\Gamma(2\mu+2\nu)}}
\left|\frac{\Gamma[\mu+\nu+{\i}(a+b)]}{\Gamma[\mu+{\i}(z+a)]\Gamma[\nu+{\i}(z-b)]}\right|,
\end{equation}
\begin{equation}\label{eq:A.5}
\delta^{\mu}(\varepsilon)
=-\arg\Gamma[\mu+{\i}(z+a)]-\arg\Gamma[\nu+{\i}(z-b)].
\end{equation}
Therefore, the scattering amplitude vanishes if $\mu+{\i}(z+a)=-k$ where
$k=0,1,\ldots,N$ and $N$ is the largest integer less than or equal to $-\mu$.
Consequently, the bound states and/or resonance energies are obtained from the following spectrum formula
\begin{equation}\label{eq:A.6}
(z+a)^2=-(k+\mu)^2.
\end{equation}
Now, the corresponding discrete wavefunction will be written in terms of the
hypergeometric function
${}_{3}F_{2}\left(\left.
-n,n+2\mu+2\nu-1,-k\atop
2\mu, \mu+\nu+{\i}(a+b)
\right|1\right)$,
which is obtained from (\ref{eq:A.1}) by the substitution $\mu+\i(z+a)=-k$.
With a suitable change of parameters
as either $2\mu=-N$, $2\nu=\alpha+\beta+N+2$, ${\i}(a+b)=\frac{\alpha-\beta}{2}$
or $2\mu=\alpha+1$, $2\nu=\beta+1$, ${\i}(a+b)=-\frac{\alpha+\beta}{2}-N-1$,
this is just the Hahn polynomial $Q_n^N(k;\alpha,\beta)$
whose normalized version is written as follows (see pages 204-208 in Ref.~\cite{koekoek2010hypergeometric}):
\begin{equation}\label{eq:A.7}
\footnotesize{
Q_{n}^{N}\left(k;\alpha,\beta\right)
=\sqrt{\left(\frac{2n+\alpha+\beta+1}{\alpha+\beta+1}\right)
\frac{(\alpha+1)_n(\alpha+\beta+1)_{N+1}(N-n+1)_n}
{(\beta+1)_n(n+\alpha+\beta+1)_{N+1}n!}}\,
{{}_{3}F_{2}}\left(\left.{-n,-k,n+\alpha+\beta+1 \atop \alpha+1,-N}\right|1\right),
}
\end{equation}
where parameter $\alpha$ or $\beta$ are either greater than $-1$ or less than $-N$.
The normalized discrete weight function is
\begin{equation}\label{eq:A.8}
\omega_k^{N}(\alpha,\beta)
=\frac{N!}{(\alpha+\beta+2)_N}\frac{(\alpha+1)_k(\beta+1)_{N-k}}{k!(N-k)!}.
\end{equation}
That is, $\sum_{k=0}^N \omega_k^{N}(\alpha,\beta) Q_n^{N}(k;\alpha,\beta)Q_m^{N}(k;\alpha,\beta)=\delta_{n,m}$.
It also satisfies the dual orthogonality
$\sum_{k=0}^N  Q_k^{N}(n;\alpha,\beta)Q_k^{N}(m;\alpha,\beta)=\delta_{n,m}/\omega_n^{N}(\alpha,\beta)$.
The symmetric three-term recursion relation satisfied by this polynomial is
obtained from that of the continuous Hahn polynomial (\ref{eq:A.3}) by the parameters map as follows
{\footnotesize
\begin{align}
&kQ_{n}^{N}
=\frac{1}{2n+\alpha+\beta+1}
\left[\frac{(N-n)(n+\alpha+1)(n+\alpha+\beta+1)}{2n+\alpha+\beta+2}
+\frac{n(n+\beta)(n+N+\alpha+\beta+1)}{2n+\alpha+\beta}\right]Q_{n}^{N}
\nonumber
\\
&-\frac{1}{2n+\alpha+\beta+2}
\sqrt{\frac{(n+1)(N-n)(n+\alpha+1)(n+\beta+1)(n+\alpha+\beta+1)(n+N+\alpha+\beta+2)}
{(2n+\alpha+\beta+1)(2n+\alpha+\beta+3)}}Q_{n+1}^{N}
\nonumber
\\
&-\frac{1}{2n+\alpha+\beta}
\sqrt{\frac{n(N-n+1)(n+\alpha)(n+\beta)(n+\alpha+\beta)(n+N+\alpha+\beta+1)}
{(2n+\alpha+\beta-1)(2n+\alpha+\beta+1)}}Q_{n-1}^{N}.
\label{eq:A.9}
\end{align}
}

\section{Asymptotics of the continuous Hahn polynomial}
\label{sec:B}

Using the method of uniform asymptotic expansion for difference equations~\cite{wang2003asymptotic,wang2005linear}
and the matching technique in the complex plane developed in~\cite{dai2014plancherel,wang2014plancherel},
the authors of \cite{cao2017aysmptotic} derived the large-$n$ asymptotic for the continuous Hahn polynomials
and their zeros via their three-term recursion relation.
Starting with the traditional definition of the polynomial $p_n(x;a,b,c,d)$
(see Eq.~(9.4.1) in \cite{koekoek2010hypergeometric}),
the authors obtained the following asymptotic formula (as $n\to\infty$)
\begin{align}
p_n\left(nt\right)
\approx
&2^{\frac52-2u}\left(\frac{n}{4e}\right)^n\frac{t^{1-u}}{(1-4t^2)^{1/4}}\exp\left\{(2nt+v)\frac\pi2\right\}
\nonumber
\\
&\times
\left\{\cos\left\{(2nt+v)\log\left[\frac{1+(1-4t^2)^{1/2}}{2t}\right]\right.\right.
\label{eq:B.1}
\\
&\hspace{12mm}\left.\left.+{\i}\Big(n+u-\frac12\Big)\log\left[2t+{\i}(1-4t^2)^{1/2}\right]+\frac{\pi}{4}\right\}
+O\left(\frac{1}{n}\right)\right\},
\nonumber
\end{align}
where $p_n(x)=\frac{n!p_n(x;a,b,c,d)}{(n+a+b+c+d-1)_n}$
is the monic version of the of the continuous Hahn polynomial,
$u=\Re(a+b)$ and $v=\Im(a+b)$.
The polynomial parameters $a$, $b$, $c$, and $d$ are complex conjugate pairs with positive real parts.
That is, $c=\bar{a}$,  $d=\bar{b}$ and $\Re a>0$, $\Re b>0$.
We start by rewriting Eq.~(\ref{eq:B.1}) at $x=nt-\frac{v}{2}$ instead of $x=nt$ giving
\begin{align}
p_n\left(nt-\frac{v}{2}\right)
\approx
&2^{\frac52-2u}\left(\frac{n}{4e}\right)^n\frac{t^{1-u}}{(1-4t^2)^{1/4}}\exp\left\{(2nt)\frac\pi2\right\}
\nonumber
\\
&\times
\left\{\cos\left\{(2nt)\log\left[\frac{1+(1-4t^2)^{1/2}}{2t}\right]\right.\right.
\label{eq:B.2}
\\
&\hspace{12mm}\left.\left.+{\i}\Big(n+u-\frac12\Big)\log\left[2t+{\i}(1-4t^2)^{1/2}\right]+\frac{\pi}{4}\right\}
+O\left(\frac{1}{n}\right)\right\}.
\nonumber
\end{align}
By setting $A(z)=\Gamma(a+{\i}z)\Gamma(b+{\i}z)$
and with the use of Stirling's formula, we have
\begin{equation}\label{eq:B.3}
A(z)\approx 2\pi \exp\big[(a+b+2{\i}z-1)\log({\i}z)-2{\i}z\big],\qquad\text{as}\quad z\to\infty.
\end{equation}
A combination of (\ref{eq:B.2}) and (\ref{eq:B.3}) gives
\begin{equation}\label{eq:B.4}
p_n(z)
\approx
2^{\frac52-2u}\pi\left(\frac{n}{4e}\right)^n n^{u-1}
\left\{\frac{1}{|A(z)|}
\cos\left[ (2z+v)\log n-\arg A(z)
-\frac{n\pi}{2}\right]
+O\left(\frac{1}{n}\right)
\right\}.
\end{equation}
To obtain the asymptotic formula for the normalized continuous Hahn polynomial (\ref{eq:A.1}),
we reparameterize the polynomial using real numbers as $a\to \mu+{\i}a$ and $b\to\nu-{\i}b$ with
$\mu$ and $\nu$ positive.
Using Eqs. (9.4.1) and (9.4.4) in~\cite{koekoek2010hypergeometric} and
the formula for the normalized continuous Hahn polynomial (\ref{eq:A.1}),
we have $P_n^{\mu}(z;\nu,a,b)=k_np_n(z)$, where $k_n$ is the leading coefficient.
With a use of Stirling's formula, we have
\begin{equation}\label{eq:B.5}
k_n\approx
\frac{1}{2\pi}\left(\frac{4e}{n}\right)^n n^{\frac12-u}
2^{2u-1}\big|\Gamma[\mu+\nu+{\i}(a+b)]\big|
\sqrt{\frac{\Gamma(2\mu)\Gamma(2\nu)}{\Gamma(2\mu+2\nu)}}.
\end{equation}
Therefore, the asymptotic formula (\ref{eq:B.4}) can be rewritten as follows
\begin{equation}\label{eq:B.6}
\begin{aligned}
P_n^{\mu}(z;\nu;a,b)
\approx
&2\sqrt{2}\big|\Gamma[\mu+\nu+{\i}(a+b)]\big|
\sqrt{\frac{\Gamma(2\mu)\Gamma(2\nu)}{\Gamma(2\mu+2\nu)}}
\times
\\
&\frac{1}{\sqrt{n}}
\left\{\frac{1}{|A(z)|}\cos\left[(2z+a-b)\log n-\arg A(z)-n\frac{\pi}{2}\right]
+O\left(\frac{1}{n}\right)\right\},
\end{aligned}
\end{equation}
where $A(z)=\Gamma[\mu+{\i}(z+a)]\Gamma[\nu+{\i}(z-b)]$ after the reparametrization.

\bibliography{HQS}

\end{document}